\documentclass[aps,prd,superscriptaddress,showpacs,preprintnumbers,10pt]{revtex4}
\usepackage{graphicx,color}
\newcommand{\be}{\begin{equation}}
\newcommand{\ee}{\end{equation}}
\newcommand{\bea}{\begin{eqnarray}}
\newcommand{\eea}{\end{eqnarray}}

\def\lsim{\mathrel{\rlap{\lower4pt\hbox{\hskip1pt$\sim$}}\raise1pt\hbox{$<$}}}
\def\gsim{\mathrel{\rlap{\lower4pt\hbox{\hskip1pt$\sim$}}\raise1pt\hbox{$>$}}}
\def\nostrocostruttino#1\over#2{\mathrel{\mathop{\kern 0pt \rlap
{\hbox{$#1$}}} \hbox{\kern-.135em $#2$}}}

\def\Vec#1{{\bf #1}}

\def\D{{\mathrm d}}
\def\E{{\mathrm e}}

\begin{document}

\title{The Boer-Mulders effect in unpolarized SIDIS: \\
an analysis of the COMPASS and HERMES data \\ on the $\cos 2 \phi$ asymmetry}

\author{Vincenzo Barone}
\affiliation{Di.S.T.A., Universit\`a del Piemonte Orientale
``A. Avogadro'', \\ and INFN, Gruppo Collegato di Alessandria, 15121
Alessandria, Italy}
\author{Stefano Melis}
\affiliation{Di.S.T.A., Universit\`a del Piemonte Orientale
``A. Avogadro'', \\ and INFN, Gruppo Collegato di Alessandria, 15121
Alessandria, Italy}
\author{Alexei Prokudin}
\affiliation{Jefferson Laboratory, 12000 Jefferson Avenue, Newport News, VA 23606}


\begin{abstract}

We present a phenomenological  analysis of the $\cos 2 \phi$ asymmetry 
recently measured by the COMPASS and HERMES collaborations
in unpolarized semi-inclusive deep inelastic scattering.  
In the kinematical regimes explored by these experiments 
the asymmetry arises from transverse-spin and intrinsic 
transverse-momentum effects. We consider 
the leading-twist contribution, related to the 
so-called Boer-Mulders transverse-polarization 
distribution $h_1^{\perp}(x, k_T^2)$, and the 
twist-4 Cahn contribution,   involving unpolarized transverse-momentum 
distribution functions.
We show that a reasonably good fit of the preliminary data sets from COMPASS and HERMES is achieved 
with a Boer-Mulders function consistent with the 
main theoretical expectations. 
Our conclusion is that the COMPASS and HERMES  
measurements represent the first experimental 
evidence of the Boer-Mulders effect in SIDIS. 

\end{abstract}

\pacs{13.88.+e, 13.60.-r, 13.85.Ni}

\maketitle

\section{Introduction} 

Among the various observables related to the transverse 
momentum and the transverse spin of quarks (for 
reviews, see \cite{Barone:2001sp,D'Alesio:2007jt}),  
the azimuthal asymmetries 
in unpolarized semi-inclusive deep inelastic scattering (SIDIS)
at moderate $P_T$ 
have recently attracted a large experimental and theoretical 
attention (see, for instance, 
\cite{Oganesian:1997jq,Gamberg:2003ey,Anselmino:2005nn,Barone:2005kt,Gamberg:2007wm,Zhang:2008ez,Burkardt:2007xm,Bacchetta:2008af,Courtoy:2009pc,Pasquini:2010af}).   
 In particular, the $\cos 2 \phi$ asymmetry potentially 
represents  a fundamental source of information on the Boer-Mulders 
function $h_1^{\perp}(x, k_T)$ \cite{Boer:1997nt},  
which measures the transverse polarization asymmetry of quarks 
inside an unpolarized hadron.   

In a previous paper \cite{Barone:2008tn}, which predated 
the experimental measurements, we presented a systematic study 
of the $\cos 2 \phi$ asymmetry  
in unpolarized SIDIS,  
taking into account both 
non perturbative and perturbative contributions. 
We found that $\langle \cos 2 \phi \rangle$
is of the order of few percent, but  
our main prediction was that, 
due to the Boer-Mulders effect, the  
asymmetry in $\pi^-$ production should be larger than in 
$\pi^+$ production. The recently released COMPASS 
\cite{Kafer:2008ud,Bressan:2009eu}
and HERMES data \cite{Giordano:2009hi} confirm this 
expectation (the $\cos 2 \phi$ asymmetry was previously 
investigated at high $Q^2$, where it is dominated by 
perturbative QCD effects \cite{Barone:2008tn}, by the 
EMC  \cite{Arneodo:1986cf} and  ZEUS \cite{Breitweg:2000qh} experiments). 

The purpose of the present work is to perform 
a phenomenological analysis of the 
recent $\langle \cos 2 \phi \rangle$
 measurements in order to check their mutual compatibility 
and to extract some information about the Boer-Mulders function. 
The available data do not allow yet a full fit 
of $h_1^{\perp}$, with its $x$ 
and $k_T$ dependence: so we will simply   
relate $h_1^{\perp}$ to its chiral-even partner, 
the Sivers function 
$f_{1T}^{\perp}$ (describing the distribution of 
unpolarized quarks in a transversely polarized 
hadron \cite{Sivers:1989cc,Sivers:1990fh}) and 
limit ourselves to fitting the proportionality 
coefficient between the two distributions. 

We will see that a reasonably good description 
of both HERMES and COMPASS preliminary data is achieved with 
a Boer-Mulders function close to 
that used in Ref.~\cite{Barone:2008tn} and 
consistent with various theoretical 
expectations (impact-parameter approach \cite{Burkardt:2005hp}, 
 lattice results \cite{Gockeler:2006zu}, 
large-$N_c$ predictions\cite{Pobylitsa:2003ty} and model 
calculations \cite{Gamberg:2007wm,Bacchetta:2008af,Courtoy:2009pc,Pasquini:2010af}.  
We found however that the quality of the fit 
depends on the assumptions about the average 
transverse momenta of quarks for each measurement.  
One problem emerges also quite clearly: while the $x$ and $z$ 
dependencies of the two sets of data are
satisfactorily reproduced,  the $P_T$ behavior  
of the COMPASS data appears to be 
incompatible with the HERMES corresponding behavior, 
and hard to understand phenomenologically.

\section{The $\cos 2 \phi$ asymmetry in unpolarized SIDIS}
\label{theory}

\begin{figure}[t]
\includegraphics[width=0.75\textwidth]
{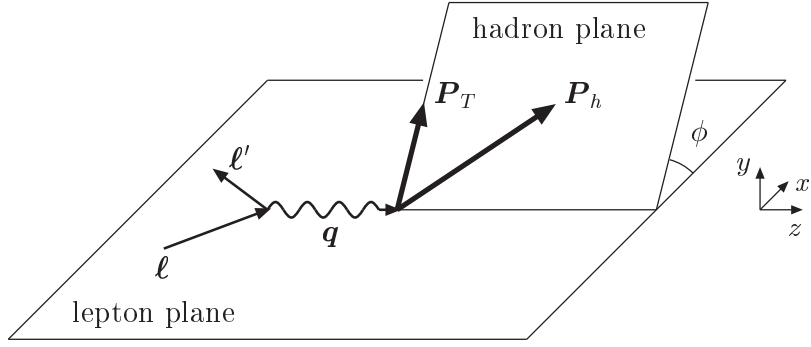}
\caption{\label{plane} Lepton and hadron planes in semi-inclusive 
deep inelastic scattering.
}
\end{figure}

The process we will consider  is unpolarized SIDIS:
\begin{equation}
l (\ell) \, + \, p (P) \, \rightarrow \, l' (\ell')
\, + \, h (P_h) \, + \, X (P_X)\,.
\label{sidis}
\end{equation}
The SIDIS cross section is expressed in terms of
the invariants
\begin{equation}
x = \frac{Q^2}{2 \, P \cdot q}, \;\;\;
y =  \frac{P \cdot q}{P \cdot \ell} ,
\;\;\;
z = \frac{P \cdot P_h}{P \cdot q}\,,
\end{equation}
where $ q = \ell - \ell'$ and $Q^2 \equiv - q^2$.
The reference frame we adopt is such that
the virtual photon and the target proton are collinear
and directed along the $z$ axis, with the 
photon moving in the positive $z$ direction 
(Fig.~\ref{plane}). We denote by $\Vec k_T$ the transverse
momentum of the quark inside the proton, by $\Vec P_T$ the
transverse momentum of the hadron $h$, and by $\Vec p_T$ the 
transverse momentum
of $h$ with respect to the direction of the fragmenting
quark. All azimuthal angles
are referred to the lepton scattering plane
(we call $\phi$ the azimuthal angle of the hadron $h$, 
see Fig.~\ref{plane}). We follow the conventions of 
Ref.~\cite{Bacchetta:2004jz}.

We are interested in the low-$P_T$ region, 
$P_T \lsim 1$ GeV. 
In Ref.~\cite{Barone:2008tn} we showed that the perturbative 
contribution to $\langle \cos 2 \phi \rangle$
 \cite{Georgi:1977tv,Mendez:1978zx,Konig:1982uk,Chay:1991nh}
is negligible in this domain.  
Therefore, in the present analysis we only take into 
account the non-perturbative effects, related to 
the transverse momentum of quarks. For small transverse momenta 
it has been formally proven that  
SIDIS can be described in terms of 
Transverse Momentum Dependent (TMD)
distribution and fragmentation functions
 \cite{Ji:2004xq}.  We will work 
at tree level, that is in the parton model generalized 
to include transverse momenta of quarks.
At leading twist, there is just one contribution,  
involving the Boer--Mulders distribution $h_1^{\perp}$
coupled to the Collins fragmentation function 
$H_1^{\perp}$ \cite{Collins:1992kk}, 
which describes the fragmentation 
of transversely polarized quarks into polarized
hadrons.  
Concerning higher-twist terms, it is known that 
there is no twist-3 contribution to $\langle \cos 2\phi \rangle$
\cite{Bacchetta:2006tn}, whereas at twist-4 (i.e.,  
at order $k_T^2/Q^2$) the asymmetry 
receives a contribution from the so-called Cahn term 
\cite{Cahn:1978se,Cahn:1989yf}, arising from the non-collinear 
kinematics of quarks.

 The 
$\phi$--independent part of the 
SIDIS differential cross section reads
\begin{eqnarray}
& &  \frac{\D^5 \sigma^{(0)}_{\rm sym}}{\D x \, \D y \, \D z \, \D^2 \Vec P_T}
 =  \frac{2 \pi \alpha_{\rm em}^2 s}{Q^4} \, \sum_a
e_a^2 \, x [1 + (1 - y)^2] \nonumber \\
& &  \hspace{1cm} \times \, \int \D^2 \Vec k_T \,
\int \D^2 \Vec p_T \,
\delta^2 (\Vec P_T - z \Vec k_T - \Vec p_T)
\, f_1^a (x, k_T^2) \,
D_1^a (z, p_T^2) \,,
\label{cross1}
\end{eqnarray}
where $f_1^a (x, k_T^2)$ is the unintegrated
number density of quarks of flavor $a$ and
$D_1^a (z, p_T^2)$ is the transverse-momentum
dependent fragmentation function of quark $a$ into
the final hadron. 

At leading-twist the only $k_T$-dependent term contributing 
to the $\cos 2 \phi$
asymmetry contains the Boer-Mulders distribution $h_1^{\perp}$
coupled to the Collins fragmentation function $H_1^{\perp}$
of the produced hadron. This contribution
to the cross section is given by \cite{Boer:1997nt}
\begin{eqnarray}
& & \left. \frac{\D^5 \sigma^{(0)}_{\rm BM}}{\D x \, \D y \, 
\D z \, \D^2 \Vec P_T}
\right \vert_{\cos 2 \phi}
 =  \frac{4 \pi \alpha_{\rm em}^2 s}{Q^4} \, \sum_a
e_a^2 \, x (1 - y) \nonumber \\
& & \hspace{2cm} \times \,
 \int \D^2 \Vec k_T \, \int \D^2 \Vec p_T
\,\delta^2 (\Vec P_T - z \Vec k_T - \Vec p_T)
\nonumber \\
& & \hspace{2cm} \times \,
\frac{2 \, \Vec h \cdot \Vec k_T \,
\Vec h \cdot \Vec p_T - \Vec k_T \cdot \Vec p_T}{z  M_N M_h} \,
h_1^{\perp a} (x, k_T^2) \,
H_1^{\perp a} (z, p_T^2)\, \cos 2 \phi \,,
\label{cross3}
\end{eqnarray}
where $M_N$ is the mass of the nucleon, $M_h$ is the 
mass of the produced hadron and $\Vec h \equiv \Vec P_T/P_T$. 

 The twist-4 Cahn contribution  
has the form
\begin{eqnarray}
& & \left. \frac{\D^5 \sigma^{(0)}_{\rm C}}{\D x \, 
\D y \, \D z \, \D^2 \Vec P_T}
\right \vert_{\cos 2 \phi}
 =  \frac{8 \pi \alpha_{\rm em}^2 s}{Q^4} \, \sum_a
e_a^2 \, x (1 - y) \nonumber \\
& & \hspace{2cm} \times \,
 \int \D^2 \Vec k_T \, \int \D^2 \Vec p_T
\,\delta^2 (\Vec P_T - z \Vec k_T - \Vec p_T)
\nonumber \\
& & \hspace{2cm} \times \,
\frac{2 \, (\Vec k_T \cdot \Vec h)^2 -  k_T^2}{Q^2} \,
f_1^a (x, k_T^2) \,
D_1^a (z, p_T^2)\, \cos 2 \phi \,. 
\label{cross2}
\end{eqnarray}
Notice that the Cahn term is only a part of the total, still unknown, twist-4 contribution.

The  asymmetry determined experimentally is defined as
\begin{equation}
A^{\cos2\phi} \equiv 2\,\langle \cos 2 \phi \rangle =2\,
 \frac{\int \D \sigma \, \cos 2 \phi}{\int \D \sigma}\,.\
\label{exp_asy}
\end{equation} 
The integrations are performed over the measured
ranges of $x, y, z$, with a lower cutoff $P_T^{\rm min}$ on $P_T$, 
which represents the minimum value of $P_T$ of the detected charged particles.

Using the expressions above, the numerator and the denominator 
of (\ref{exp_asy}) are given by 
\begin{eqnarray}
& & 
\int \D \sigma \cos 2 \phi =
\frac{4 \pi \alpha_{\rm em}^2 s}{Q^4} \, 
\int \int \int \int \,
 \sum_a e_a^2 \,  x (1- y) \, \{ \mathcal{A} [f_1^a, D_1^a] +
\frac{1}{2} \, \mathcal{B} [h_1^{\perp a}, H_1^{\perp a}] \}\,, \\
& & \int \D \sigma =
\frac{2 \pi \alpha_{\rm em}^2 s}{Q^4} \, 
\int \int \int \int \,
\sum_a e_a^2 \, x   [1 +
(1 - y)^2]  \, \mathcal{C} [f_1^a,
D_1^a]\,, 
\end{eqnarray}
where
\begin{equation}
\int \int \int \int \equiv
\int_{P_T^{\rm min}}^{P_{T}^{\rm max}} \D P_T \, P_T
\, \int_{x_1}^{x_2} \D x \,
\int_{y_1}^{y_2} \D y \, \int_{z_1}^{z_2} \D z \,
\end{equation}
and
($\chi$ is the angle between $\Vec P_T$
and $\Vec k_T$)
\begin{eqnarray}
\mathcal{A} [f_1^{a}, D_1^{a}] &\equiv&
 \int \D^2 \Vec k_T \, \int \D^2 \Vec p_T
\,\delta^2 (\Vec P_T - z \Vec k_T - \Vec p_T)
\nonumber \\
& &  \times \,
\frac{2 \, (\Vec k_T \cdot \Vec h)^2 -  k_T^2}{Q^2} \,
f_1^a (x,  k_T^2) \,
D_1^a (z,  p_T^2)\, \cos 2 \phi
\nonumber \\
& =&
\int_0^{\infty} \D k_T \, k_T \, \int_0^{2 \pi} \D \chi
\, \frac{2 \, k_T^2 \, \cos^2 \chi -  k_T^2}{Q^2}
\nonumber \\
& & \times \,  f_1^{a}(x,  k_T^2)
\, D_1^{a}(z, \vert \Vec P_T - z \Vec k_T \vert^2)\,,
\label{convol1}
\end{eqnarray}
\begin{eqnarray}
\mathcal{B} [h_1^{\perp a}, H_1^{\perp a}] &\equiv&
 \int \D^2 \Vec k_T \, \int \D^2 \Vec p_T
\,\delta^2 (\Vec P_T - z \Vec k_T - \Vec p_T)
\nonumber \\
& &  \times \,
\frac{2 \, \Vec h \cdot \Vec k_T \,
\Vec h \cdot \Vec p_T - \Vec k_T \cdot \Vec p_T}{z  M_N M_h} \,
h_1^{\perp a} (x,  k_T^2) \,
H_1^{\perp a} (z,  p_T^2)
\nonumber \\
& =&
\int_0^{\infty} \D k_T \, k_T \, \int_0^{2 \pi} \D \chi
\, \frac{ k_T^2 + (P_T/z)\, k_T
\, \cos \chi - 2 \,  k_T^2 \, \cos^2 \chi}{M M_h}
\nonumber \\
& & \times \, h_1^{\perp a}(x,  k_T^2)
\, H_1^{\perp a}(z, \vert \Vec P_T - z \Vec k_T \vert^2)\,,
\label{convol2}
\end{eqnarray}
\begin{eqnarray}
\mathcal{C} [f_1^{a}, D_1^{a}] &\equiv&
\int \D^2 \Vec k_T \,
\int \D^2 \Vec p_T \,
\delta^2 (\Vec P_T - z \Vec k_T - \Vec p_T)
\, f_1^a (x,  k_T^2) \,
D_1^a (z,  p_T^2)
\nonumber \\
& =&
\int_0^{\infty} \D k_T \, k_T \, \int_0^{2 \pi} \D \chi
\,  f_1^{a}(x,  k_T^2)
\, D_1^{a}(z, \vert \Vec P_T - z \Vec k_T \vert^2)\,.
\label{convol3}
\end{eqnarray}

\section{Parametrizations of distribution and fragmentation functions}

To calculate the azimuthal asymmetries we need first of all 
the $k_T$-dependent unpolarized distribution functions, 
which we assume to have a Gaussian behavior in $k_T$: 
\be
f_1^q (x, k_T^2) = f_1^q (x) \frac{\E^{-{ k_T^2}/
{\langle  k_T^2 \rangle}}}{\pi \langle  k_T^2 \rangle }\,. 
\ee
The Gaussian dependence of the transverse-momentum 
distribution functions
is supported by a recent lattice study \cite{Musch:2007ya}. 
The integrated unpolarized distribution 
functions $f_1^q$ are taken
from the GRV98 fit \cite{Gluck:1998xa}.

The available data on $\langle \cos 2 \phi \rangle$ 
do not allow  a full extraction of the Boer-Mulders function. 
 Thus we simply take $h_1^{\perp}$ to be proportional 
to the Sivers function $f_{1T}^{\perp}$, 
 \be
h_1^{\perp q}(x, k_T^2) = \lambda_q \,  
f_{1T}^{\perp q}(x, k_T^2)\,, 
\label{lambda}
\ee 
with a coefficient $\lambda_q$ to be fitted to the data. 
Various theoretical arguments (based 
on the impact-parameter picture \cite{Burkardt:2005hp}, on 
large-$N_c$ arguments \cite{Pobylitsa:2003ty}, and on model calculations 
\cite{Bacchetta:2008af,Courtoy:2009pc,Pasquini:2010af}) 
suggest that  that the $u$ and $d$ components 
of $h_1^{\perp}$, at variance with $f_{1T}^{\perp}$, 
 should have the same sign and in particular 
be both negative (which means that $\lambda_d$ should be 
negative). This is indeed what we find in our analysis. 
Moreover, the impact-parameter approach \cite{Burkardt:2005hp} combined 
with lattice results \cite{Gockeler:2006zu} predicts 
a $u$ component of $h_1^{\perp}$ larger in magnitude than 
the corresponding component of $f_{1T}^{\perp}$, and the  
$d$ components of $h_1^{\perp}$ and $f_{1T}^{\perp}$ with approximately 
the same magnitude (and  
opposite sign). 

We parametrize the Boer--Mulders function using the 
Ansatz (\ref{lambda}) and taking 
the Sivers function from a fit to 
single--spin asymmetry 
data \cite{Anselmino:2008sga}. Thus we set
\be
h_1^{\perp q} (x, k_T^2)= 
\lambda_q \, f_{1T}^{\perp q} (x, k_T^2)=  
\lambda_q \, \rho_q(x) \, \eta(k_T) \, 
f_1^q(x, \Vec k_T^2) \, , \label{sivfac}
\ee
where
\bea
&&\rho_q(x) = A_q \, x^{a_q}(1-x)^{b_q} \,
\frac{(a_q+b_q)^{(a_q+b_q)}}{a_q^{a_q} b_q^{b_q}}\; , 
\label{siversx} \\
&&\eta (k_T) = \sqrt{2e}\,\frac{ M_P}{M_1}\,\E^{-k_T^2/M_1^2}\; \cdot
\label{siverskt}
\eea
Here $M_P$ is the proton mass, $A_q$, $a_q$, $b_q$ and $M_1$ are parameters determined in 
\cite{Anselmino:2008sga} (see Table~\ref{fitpar}). 
Being a quark spin asymmetry, 
$f_{1T}^{\perp}$ must satisfy a positivity bound, which is automatically fulfilled 
by the parametrization of Ref.~\cite{Anselmino:2008sga}. 
Notice that the Sivers function parametrization, as defined in Ref.~\cite{Anselmino:2008sga}, is: $\Delta^N\! f_q(x,k_{\perp})=-2\frac{k_{\perp}}{M_P}\>f_{1T}^{\perp q}(x,k_{\perp})$.

\begin{table}[t]
\begin{center}
\begin{tabular}{|l l l|}
\hline
~&~&~\\
~~~$A_{u} = -0.35  $ \hspace*{1cm} &
~~~$A_{d} = 0.90  $ &
~~~$A_{s} = 0.24  $~~ \\
~~~$A_{\bar u} =  -0.04  $&
~~~$A_{\bar d} =  0.40 $&
~~~$A_{\bar s} =  -1 $ \\
~~~$\alpha _u = 0.73$ &
~~~$\alpha_d = 1.08$  &
~~~$\alpha_{sea} = 0.79 $ \\
~~~$\beta \;\;= 3.46 $ &
~~~$M_1^2 = 0.34 $ (GeV/$c)^2$~  &
~~~~~ \\
~&~&~\\
\hline
\end{tabular}
\end{center}
\caption{Parameters of the Sivers function 
used in Eqs.~(\ref{siversx},\ref{siverskt})}
\label{fitpar}
\end{table}

Concerning the antiquark Boer-Mulders distributions,  
the SIDIS (at least, the present ones) 
 are not able to constrain them. Thus we simply 
take the Boer-Mulders antiquark distributions to be equal 
in magnitude to the corresponding Sivers distributions and 
both negative.  
Note that the Drell-Yan measurements 
of the $\cos 2 \phi$ asymmetry \cite{Zhu:2006gx,Zhu:2008sj} 
would in principle 
give information about the antiquark sector \cite{Zhang:2008nu,Lu:2009ip}, but 
most of the present data seem to be explainable in terms 
of perturbative QCD \cite{Boer:2006eq,Berger:2007jw}.

Let us now turn to the fragmentation functions. 
We distinguish their favored and  
unfavored components, according to 
the following general relations
\bea
&& D_{\pi^+/u} = D_{\pi^+/\bar d} = D_{\pi^-/d} = D_{\pi^-/\bar u}
\equiv D_{\rm fav} \label{fav} \\
&& D_{\pi^+/d} = D_{\pi^+/\bar u} = D_{\pi^-/u} = D_{\pi^-/\bar d}
= D_{\pi^\pm/s} =  D_{\pi^\pm/\bar s} \equiv D_{\rm unf}, \label{unf}
\eea

The $p_T$--dependent unpolarized fragmentation $D_1 (z, p_T^2)$
is assumed to have the form  
\begin{equation} 
D_1(z, p_T^2) = D_1 (z) \, \frac{\E^{- p_T^2/\langle p_T^2 \rangle}}{\pi 
\langle p_T^2 \rangle}\,,   
\end{equation} 
again with a Gaussian behavior in $p_T$. 
The integrated 
fragmentation function $D_1(z)$ is taken from the 
the DSS fit \cite{deFlorian:2007aj}.

For the Collins function we use the parametrization 
of
\cite{Anselmino:2008jk}, based on a combined 
analysis of SIDIS and $e^+ e^-$ data: 
\begin{equation} 
H_1^{\perp q} (z, p_T^2) = 
\rho_q^C(z) \, \eta^C(p_T) \, D_1 (z, p_T^2)\,, 
\label{coll-funct}
\end{equation} 
with
\bea
&&\rho_q^C(z)= A_q^C \, z^{\gamma} (1-z)^{\delta} \,
\frac{(\gamma + \delta)^{(\gamma +\delta)}}
{\gamma^{\gamma} \delta^{\delta}} \label{NC}\\
&&
\eta^C(p_T)=
\sqrt{2 e}\,\frac{z M_h}{M_C} \, 
\E^{-{p_T^2}/{M_C^2}}\,,\label{h-funct}
\eea
We let the coefficients $A_q^C$ to be flavor dependent
($q = u,d)$, while all the exponents $\gamma, \delta$ and the
dimensional parameter $M_C$ are taken to be flavor independent.
The  parametrization is devised in such a way that the Collins function 
satisfies the positivity bound (remember that 
$H_1^{\perp}$ is essentially a transverse momentum 
asymmetry).  
The values of the parameters as determined in the fit 
of Ref.~\cite{Anselmino:2008jk} are listed in Table~\ref{fitpar1}.

\begin{table}[t]
\begin{center}
\begin{tabular}{|l|llllll|}
\hline
~&~&~&~&~&~&~\\
~~Collins &
~~$A_{fav}^C$  &=& $0.44  $ & $N_{unf}^C$   &=& $-1.00  \,\,\,$ \\
~~fragmentation~~ &
~~$\gamma$  &=& $0.96  $ & $\delta$   &=& $0.01  $    \\
~~function &
~~$M_C^2$ &=&  $0.91 $ (GeV$^2$/c) &    & &  \\
~&~&~&~&~&~&~\\
\hline
\end{tabular}
\end{center}
\caption{Parameters of the favored and unfavored Collins
fragmentation functions~\cite{Anselmino:2008jk}. 
\label{fitpar1}}
\end{table}

The remaining crucial ingredients to be considered are  
the average values of $k_T^2$ and $p_T^2$. 
Notice that the following kinematical relation holds 
between the transverse momentum of the produced hadron 
and the transverse momenta of quarks: 
\be
\langle P_T^2 \rangle = \langle p_T^2 \rangle 
+ z^2 \, \langle k_T^2 \rangle \,.
\label{momenta}
\ee

Due to the limitations of the 
present data sets it is not possible 
to treat $\langle k_T^2 \rangle$ and $\langle p_T^2 \rangle$
as two additional parameters to be determined by the fit. 
We have to make some assumptions about them.  

 In our main fit (hereafter called Fit 1) we take 
$\langle k_T^2 \rangle$ and $\langle p_T^2 \rangle$
from the 
analysis of the azimuthal dependence 
of the unpolarized SIDIS cross section 
performed in Ref.~\cite{Anselmino:2005nn}:  
\be
\langle k_T^2 \rangle = 0.25 \;\; {\rm GeV}^2\,, 
\;\;\;\;
\langle p_T^2 \rangle = 0.20 \;\; {\rm GeV}^2\,. 
\label{compass_T}
\ee

We also tried another fit 
(``Fit 2''),  using for HERMES 
the values of $\langle k_T^2 \rangle$ 
and $\langle p_T^2 \rangle$ given by their own analysis 
of the $P_T$ spectrum, which turns out to be  
reproduced by Monte Carlo calculations \cite{Giordano:2008zzc}
with $\langle k_T^2 \rangle = 0.18$ ${\rm GeV}^2$
and a  $z$-dependent transverse 
momentum of the fragmenting quark, 
$\langle p_T^2 \rangle = 0.42\, z^{0.37} (1-z)^{0.54}$ ${\rm GeV}^2$.
In the $z$ range of interest this is very well 
approximated by $\langle p_T^2 \rangle \simeq 0.20$ ${\rm GeV}^2$.
Thus in our Fit 2 we choose for HERMES:
\be
\langle k_T^2 \rangle = 0.18 \;\; {\rm GeV}^2\,, 
\;\;\;\;
\langle p_T^2 \rangle = 0.20 \;\; {\rm GeV}^2\,. 
\label{hermese_T}
\ee
We have no similar information for the COMPASS measurement
 and for their data  we still use in Fit 2 the values (\ref{compass_T}). 
Therefore Fit 2 is characterized by a  
$\langle k_T^2 \rangle$ which is different for the two sets 
of data. As we shall see in the next Section, Fit 2 
turns out to be significantly better than Fit 1.

Finally, concerning a possible flavor-dependence of 
$\langle k_T^2 \rangle$ \cite{Mkrtchyan:2007sr}, we showed in 
our previous paper \cite{Barone:2008tn} that it hardly  
 affects the results, hence we shall not take it into account 
here (it should also be remarked that the experimental 
evidence for a flavor-dependent $\langle k_T^2 \rangle$  
is still far from being established). 

In summary the assumptions of our fits are:
\begin{itemize}
 \item $h_1^{\perp q} (x, k_T^2)= 
\lambda_q \, f_{1T}^{\perp q} (x, k_T^2)$ for $u$ and $d$ quarks while $h_1^{\perp \bar{q}} (x, k_T^2)=-|f_{1T}^{\perp \bar{q}} (x, k_T^2)|$ for sea quarks, with $f_{1T}^{\perp q,\bar{q}}$ functions as given in Ref.~\cite{Anselmino:2008sga}.
\item The Collins functions $H_{1}^{\perp q}(z,p_T)$ is taken as in Ref.~\cite{Anselmino:2008jk}.
\item Gaussian transverse momentum distribution is assumed for all the distribution/fragmentation functions.
\item In particular the average transverse momenta for the unpolarized distribution and fragmentation functions are respectively:
\begin{description}
 \item[]Fit 1: $\langle k_T^2 \rangle = 0.25\;\; {\rm GeV}^2$, $\langle p_T^2 \rangle = 0.20 \;\; {\rm GeV}^2$ for both HERMES and COMPASS data.
\item[]Fit 2: $\langle k_T^2 \rangle = 0.25\;\; {\rm GeV}^2$, $\langle p_T^2 \rangle = 0.20 \;\; {\rm GeV}^2$ for COMPASS data;
\item[]\hspace{6ex} $\langle k_T^2 \rangle = 0.18 \;\; {\rm GeV}^2$,
$\langle p_T^2 \rangle = 0.20 \;\; {\rm GeV}^2$ for HERMES data.
\end{description}

\end{itemize}


\section{Analysis of the  $\langle \cos 2 \phi \rangle$ data}

Data on the $\cos 2 \phi$ asymmetry in unpolarized SIDIS 
at small $P_T$ have 
been recently presented in a preliminary 
form by both the HERMES Collaboration 
\cite{Giordano:2009hi} and by COMPASS \cite{Kafer:2008ud,Bressan:2009eu} 
(while this analysis was in progress the CLAS Collaboration at JLab 
has released some results on $\langle \cos 2 \phi 
\rangle$, but their conclusion is that 
the precision of the data does not 
allow obtaining information about the Boer-Mulders function 
\cite{Osipenko:2008rv}).   
The first qualitative evidence coming from both 
COMPASS and HERMES measurements 
is a larger asymmetry for $\pi^-$ production compared to 
$\pi^+$. This difference was predicted in Ref.~\cite{Barone:2008tn} 
to be a signature of the Boer-Mulders effect, which has opposite 
signs for $\pi^+$ and $\pi^-$ (whereas the Cahn contribution, 
is the same for $\pi^+$ and $\pi^-$).

\begin{figure}[t]
\includegraphics[width=0.5\textwidth, angle=-90]
{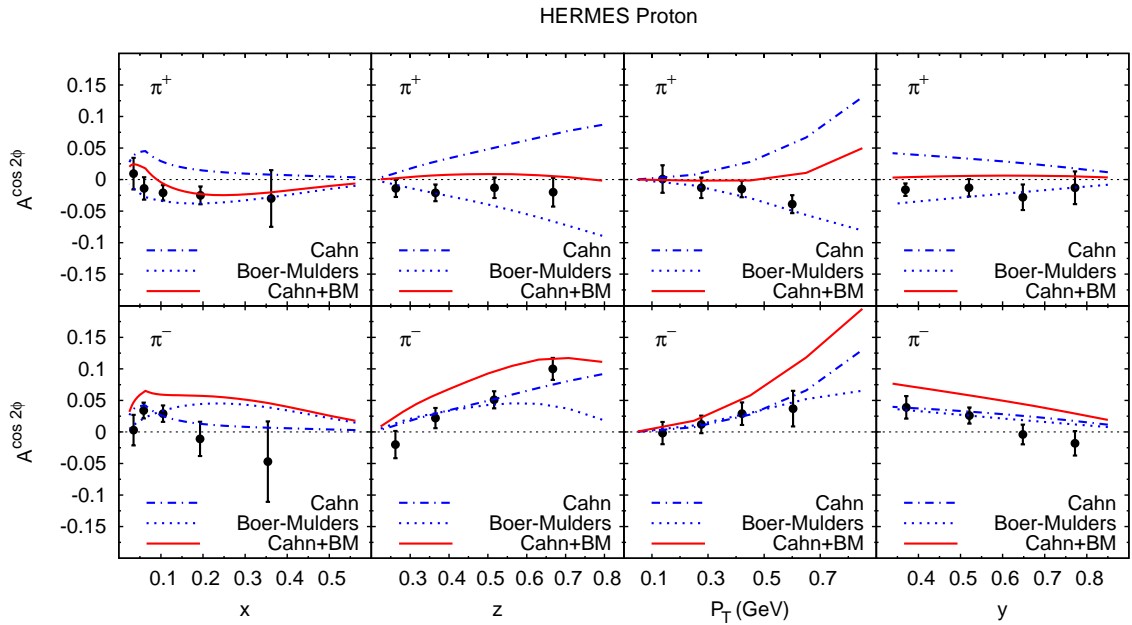}
\caption{\label{fig:hermesp_1}
Our Fit 1 to the HERMES preliminary proton-target data \cite{Giordano:2009hi}.
The dot--dashed line is  the Cahn contribution, 
the dotted line is the Boer-Mulder contribution,    
the continuous line is the resulting 
asymmetry taking both contributions into account.}
\end{figure}

Fitting the HERMES and COMPASS data as explained in the previous 
Section we find in Fit 1 
the following values for the coefficients $\lambda_u$ 
and $\lambda_d$: 
\be
\lambda_u = 2.0 \pm 0.1\,, \;\;\;\;
\lambda_d = - 1.111 \pm 0.001 \,
\;\;\;\;\;\;\; ({\rm Fit} \; 1) 
\label{lambdafit}
\ee
This implies that $h_{1}^{\perp u}$ and $h_{1}^{\perp d}$ are both negative.
The $\chi^2$  per degree of freedom of Fit 1 is 
$\chi^2/d.o.f. = 3.73$. The value of $\lambda_d$ 
corresponds to the saturation of the positivity 
bound of $h_{1}^{\perp d}$.
Errors on the parameters were calculated with $\Delta \chi^2 = 1$.

Notice that we have excluded 
from the fit the COMPASS data in $P_T$ , 
which -- as we will see below -- are clearly incompatible 
with the HERMES $P_T$ data and have a counter-intuitive 
behavior (it is in fact difficult 
to envisage a transverse-momentum dependence 
of distribution  and fragmentation functions able to describe them).

\begin{figure}[t]
\includegraphics[width=0.3\textwidth,angle=-90]
{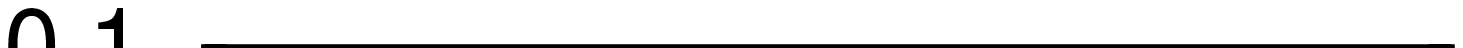}
\caption{\label{fig:bm}
The first moments of $h_{1}^{\perp u}$ and $h_{1}^{\perp d}$ 
from Fit 1.}
\end{figure}

The first moments (in $k_T^2$) of the 
Boer-Mulders distributions $h_1^{\perp u}$ 
and $h_1^{\perp d}$, 
\be
h_1^{\perp (1)}(x) \equiv 
\int \D^2 \Vec k_T \, \frac{k_T^2}{2 M^2}\, h_1^{\perp}(x, k_T^2)\,, 
\ee
are displayed in Fig.~\ref{fig:bm}.

In Fig.~\ref{fig:hermesp_1} we plot $A^{\cos 2 \phi}\equiv 2\,\langle
\cos2\phi \rangle$ for $\pi^+$ and $\pi^-$
production at HERMES with a proton target, 
with the results of Fit 1. 
The asymmetry is shown as 
a function of one 
variable at a time, $x$,
$z$ and $P_T$; the integration over the unobserved variables has been
performed over the measured ranges of the HERMES experiment, 
\bea 
 && Q^2 > 1 \; {\rm GeV}^2\,, \quad W^2 > 10 \; 
{\rm GeV}^2 \,, 
\quad P_T > 0.05 \; {\rm GeV} 
\label{hermutcuts} \\ 
&& 0.023 < x < 1.0\,, \quad 0.2 < z < 1.0 \,, \quad
0.3 < y < 0.85 
\nonumber 
\\ && 0.2 < x_F < 1
\nonumber \>. 
\eea
Note that 
the Boer-Mulders contributions 
to $\pi^+$ and $\pi^-$ production are opposite in sign. In fact, we have
\bea
\langle \cos 2 \phi \rangle^{\pi^+}_{\rm BM} \sim e_u^2 \, 
h_1^{\perp u}(x) \, H_1^{\perp {\rm fav}}(z) + e_d^2 \, h_1^{\perp d}(x) \, 
H_1^{\perp {\rm unf}}(z) \, ,\nonumber \\
\langle \cos 2 \phi \rangle^{\pi^-}_{\rm BM} \sim e_u^2 \, 
h_1^{\perp u}(x)\, H_1^{\perp {\rm unf}}(z) + e_d^2 \, h_1^{\perp d}(x) \, 
H_1^{\perp {\rm fav}}(z) \, ,
\eea
and, as far as $H_1^{\perp {\rm unf}}(z) \simeq - H_1^{\perp {\rm fav}}(z)$ 
\cite{Anselmino:2007fs}, one gets different signs for the 
Boer-Mulders effect for positive and negative pions. 
The combination of the Boer-Mulders term with the Cahn term,  
which is positive and exactly the same for $\pi^+$ and $\pi^-$ 
(if the $k_T$-dependence of the distributions is 
flavor blind) gives a resulting asymmetry which
is larger for $\pi^-$ than for $\pi^+$.

\begin{figure}[t]
\includegraphics[width=0.5\textwidth,angle=-90]
{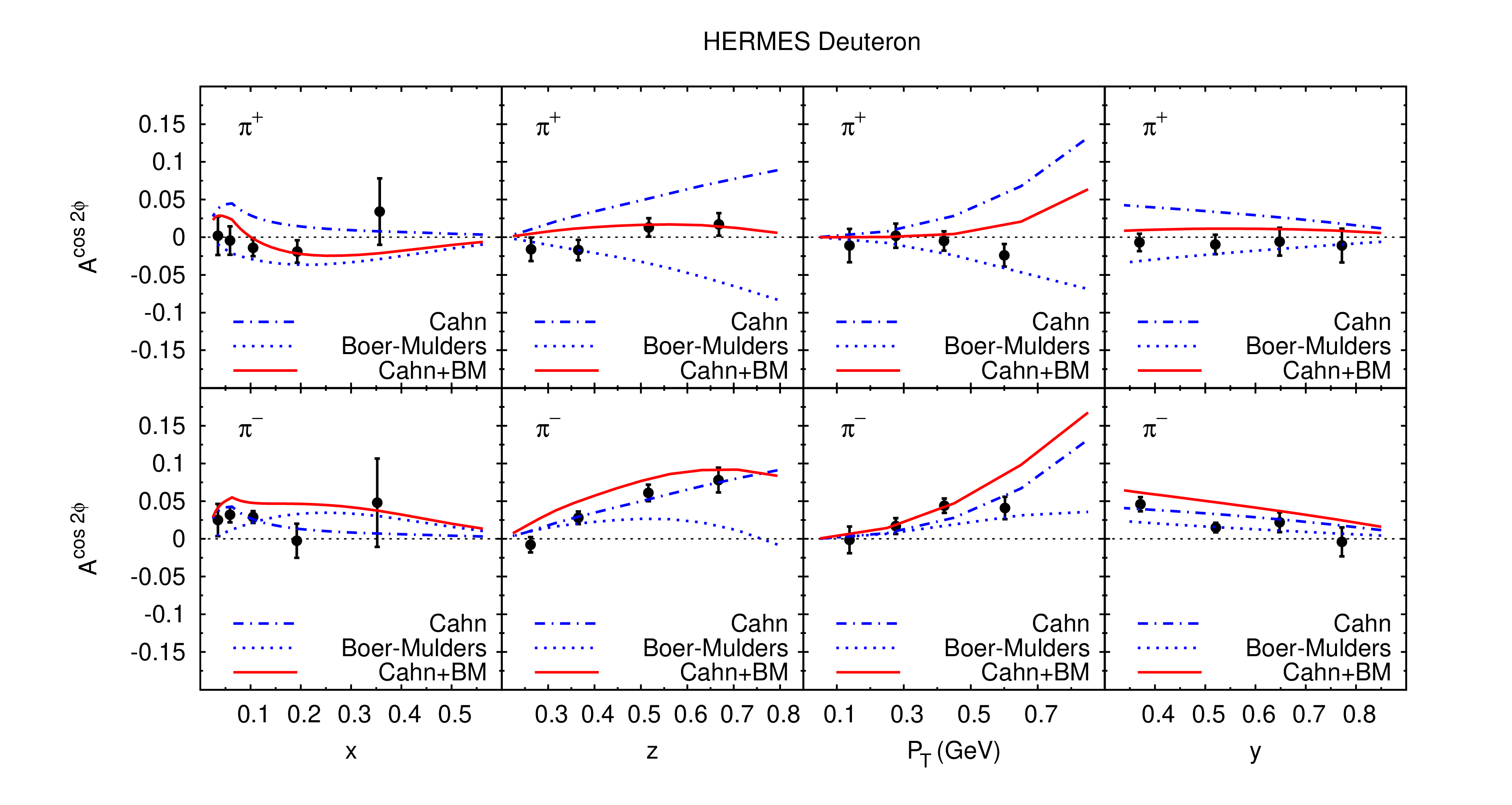}
\caption{\label{fig:hermesd_1}
Our Fit 1 to the HERMES preliminary deuteron-target data. 
The line labels are the same as in Fig.~\ref{fig:hermesp_1}.}
\end{figure}

Fig.~\ref{fig:hermesd_1} shows our Fit 1  
to $A^{\cos 2\phi_h}$ at HERMES
with a deuteron target.  
We have neglected nuclear corrections and
used isospin symmetry to relate the distribution functions of 
the neutron to those of the proton.

\begin{figure}[t]
\includegraphics[width=0.5\textwidth,angle=-90]
{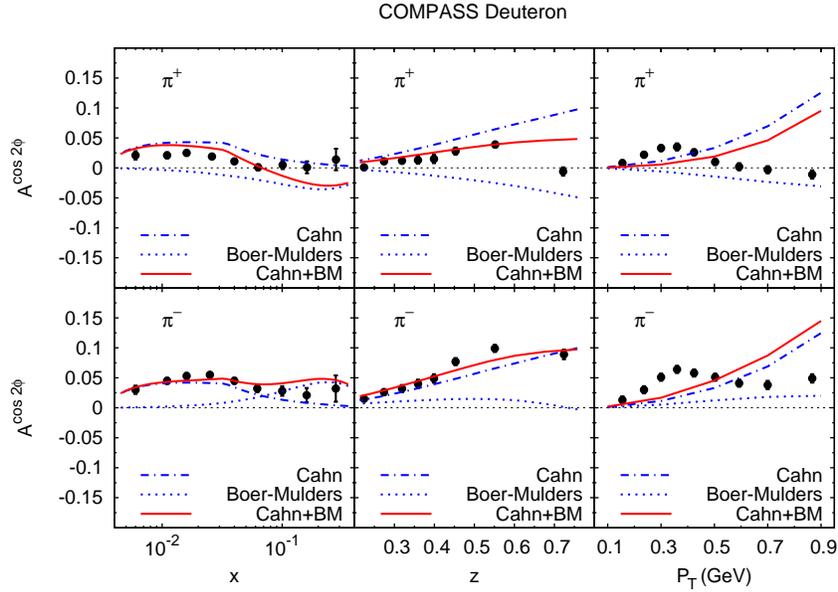}
\caption{\label{fig:compassd_1}
Our Fit 1 to the COMPASS preliminary data (deuteron target) 
\cite{Kafer:2008ud,Bressan:2009eu}. 
The line labels are the same as in Fig.~\ref{fig:hermesp_1}.}
\end{figure}

The experimental cuts of the COMPASS experiment (which 
runs with a deuteron target) are: 
\bea 
 && Q^2 \ge 1 \; {\rm GeV}^2\,, \quad W^2 > 25 \; 
{\rm GeV}^2 \,, 
\label{compasscuts} \\ 
&& 0.2 < z < 0.85 \,, \quad
0.1 \le y \le 0.9 \nonumber \\ &&  P_T > 0.1 \> {\rm GeV} \>. 
\eea
In Fig.~\ref{fig:compassd_1} we show our fit 
to the COMPASS data. One clearly sees that 
the $P_T$ dependence of these data is 
incompatible with the HERMES one and hard to understand 
theoretically.

\begin{figure}[t]
\includegraphics[width=0.5\textwidth,angle=-90]
{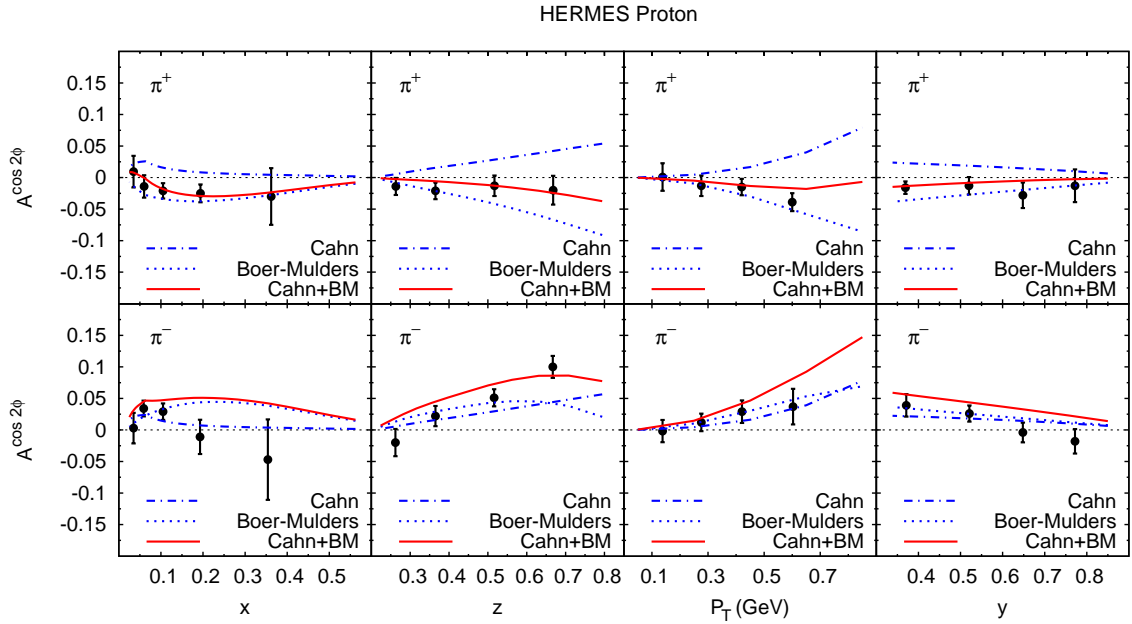}
\caption{\label{fig:hermesp_2}
Our Fit 2 to the HERMES preliminary proton data. 
The line labels are the same as in Fig.~\ref{fig:hermesp_1}.}
\end{figure}

\begin{figure}[t]
\includegraphics[width=0.5\textwidth,angle=-90]
{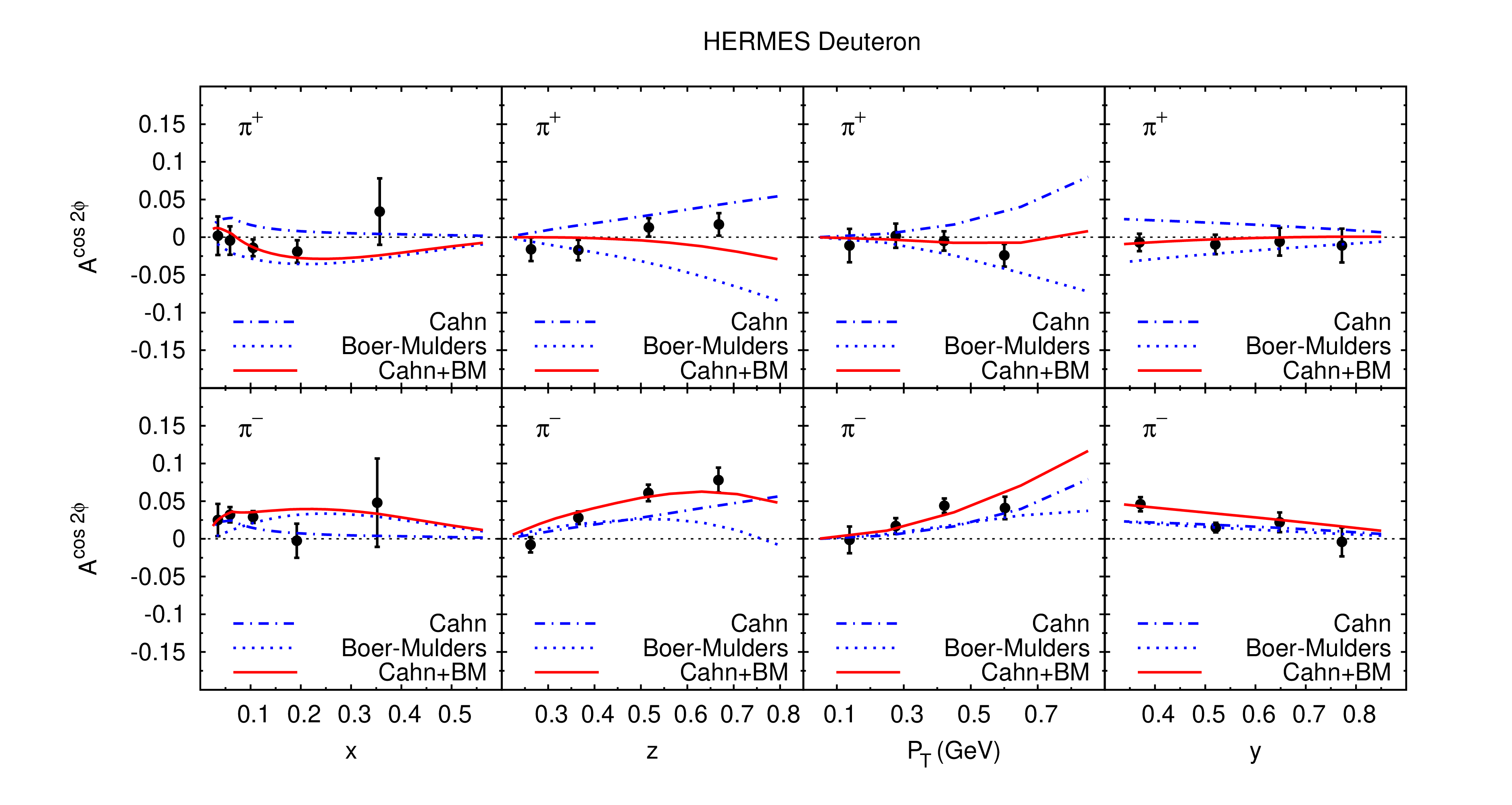}
\caption{\label{fig:hermesd_2}
Our Fit 2 to the HERMES preliminary deuteron target data. 
The line labels are the same as in Fig.~\ref{fig:hermesp_1}.}
\end{figure}

\begin{figure}[t]
\includegraphics[width=0.5\textwidth,angle=-90]
{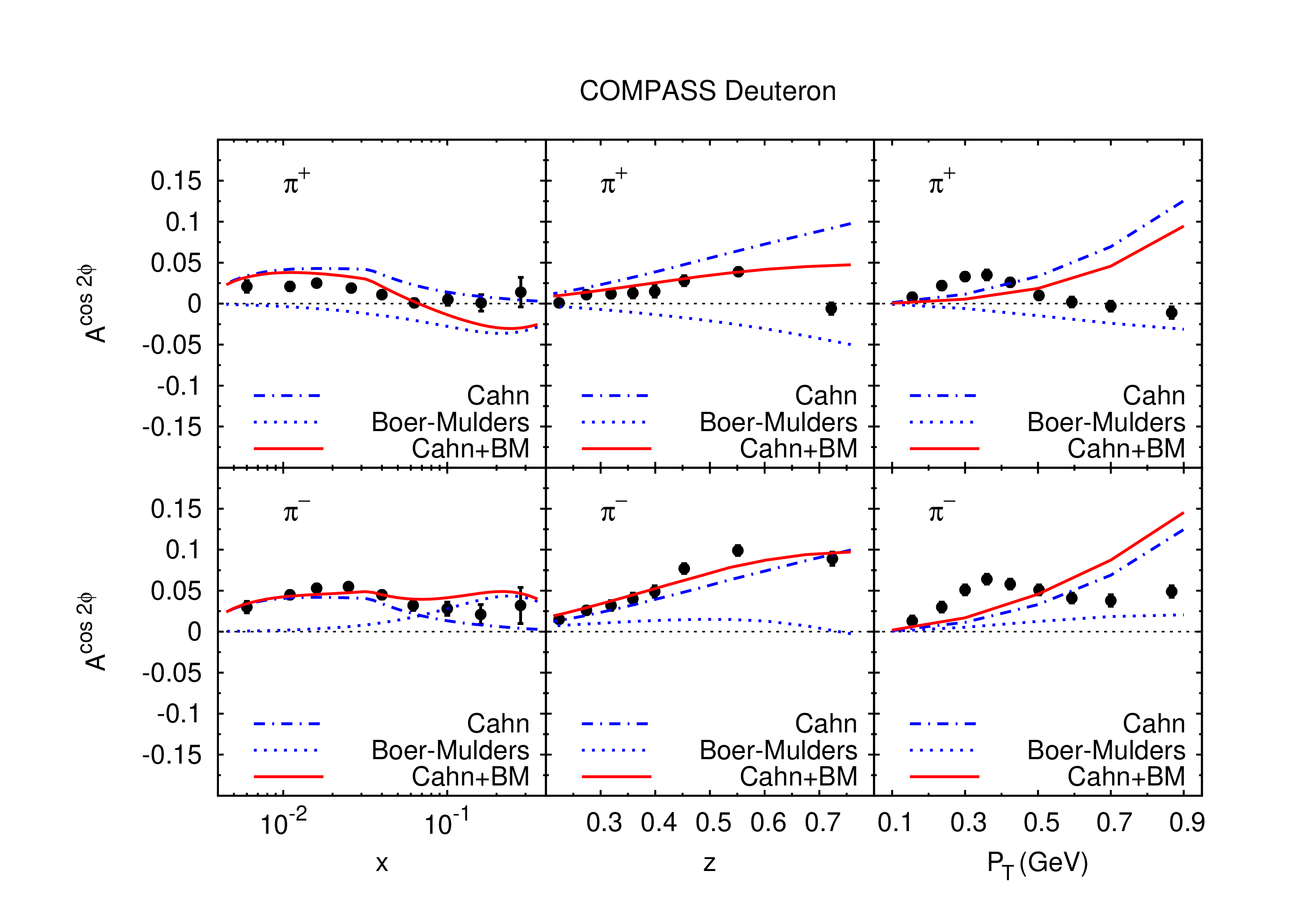}
\caption{\label{fig:compassd_2}
Our Fit 2 to the COMPASS preliminary data. 
The line labels are the same as in Fig.~\ref{fig:hermesp_1}.}
\end{figure}

Let us now come to Fit 2. 
In this case, the coefficients $\lambda_u$ 
and $\lambda_d$ are found to be  
\be
\lambda_u = 2.1 \pm 0.1\,, \;\;\;\;
\lambda_d = - 1.111 \pm 0.001 \,
\;\;\;\;\;\;\; ({\rm Fit} \; 2)\,,  
\label{lambdafit2}
\ee
and are very close to those of Fit 1 (again, $h_{1}^{\perp d}$
saturates its positivity bound). Thus, the $x$-dependence 
of the Boer-Mulders functions is essentially the 
same in the two fits. However, the  
$\chi^2$  per degree of freedom of Fit 2 is 
significantly smaller: 
$\chi^2/d.o.f. = 2.41$. In Figs.~\ref{fig:hermesp_2}, 
\ref{fig:hermesd_2} and \ref{fig:compassd_2} we show 
the results of Fit 2 for $A^{\cos 2 \phi}$ compared to the 
data. 
 
In both Fit 1 and Fit 2 we used only statistical and,
when provided (HERMES), systematic uncertainties
of the experimental data.
Given the quality of present data and the theoretical uncertainties
related to the twist four contributions,
we did not attempt a more sophisticated analysis
of the $x$ dependence of Boer-Mulders functions
and used the simplified assumption of Eq.~\ref{lambda}.

The main difference between the two fits resides 
in the Cahn term, which is strongly sensitive to 
the average value of $k_T^2$. The fact that the data 
prefer the fit with $\langle k_T^2 \rangle_{\rm HERMES} 
\neq \langle k_T^2 \rangle_{\rm COMPASS}$ 
 seems to indicate that the twist-4 contributions 
are different in the kinematics of the two  
experiments. This is a further indication of the fact that
the $\cos 2\phi$ fit is strongly affected
by twist-4 contributions, which are not yet fully known.

Our analysis shows that,  
as far as the $x$ and 
$z$ dependencies are concerned, both the HERMES and COMPASS 
data are fairly well described. The resulting Boer-Mulders distributions 
of quarks have the expected sign 
\cite{Pobylitsa:2003ty, Pasquini:2008ax,Courtoy:2009pc}. 
Moreover, looking at eq.~(\ref{lambdafit}) or (\ref{lambdafit2}), 
one sees that the Boer-Mulders $u$ distribution is larger by 
a factor 2 compared to the $u$ Sivers distribution, whereas 
the Boer-Mulders and Sivers $d$ distributions have 
approximately the same magnitude. This is in agreement 
with the predictions of the impact-parameter 
approach \cite{Burkardt:2005hp} combined with lattice results 
\cite{Gockeler:2006zu}.

\section{Conclusions and perspectives}

The present unpolarized SIDIS data 
on azimuthal $\cos 2\phi$ asymmetries,
although still preliminaries,
represent a clear manifestation  of the 
Boer-Mulders effect.  However, they are not sufficient 
to allow a full extraction of $h_1^{\perp}$.
In particular, twist-4 Cahn contribution turns out to be comparable in size
to twist-2 Boer-Mulders contribution
so measurements of the $\cos 2 \phi$ asymmetry
at different values of $Q^2$ are desirable.
Measurements at proposed Electron Ion Collider \cite{Deshpande:2005wd,Horn:2009cu} at higher $Q^2$
would allow separation of pure twist-2 from higher twist contributions.
To minimize the influence of Cahn contribution
which is supposed to be flavor blind
one can measure $\cos 2 \phi$ asymmetry of the difference $\pi^- - \pi^+$. 
Moreover, the antiquark Boer-Mulders distributions turn out to be 
essentially unconstrained. Therefore, it would 
be important not only to have more SIDIS data 
(the JLab experiments will play in the next future 
a very important role in this respect), but also 
to explore other processes, like Drell-Yan (DY) production. 
The kinematics 
of the  recent E866/NuSea DY data \cite{Zhu:2006gx,Zhu:2008sj}
is such that they are partly dominated by perturbative effects.
An attempt to extract $h_1^{\perp}$ from these data has been made in 
Refs.~\cite{Zhang:2008nu,Lu:2009ip}.
The first moments of quark Boer-Mulders functions found in these analysis agree with our results in the relative sign  of $u$ and $d$ quark Boer-Mulders functions.
The magnitudes cannot be easily compared.
First of all, as we said,
the extraction of $h_{1}^{\perp}$ from E866/NuSea data is affected
by perturbative effects which
in Refs.~\cite{Zhang:2008nu,Lu:2009ip},
have not be taken into account even at large $q_T$.
Second
in $pp$ and $pD$ DY  processes it is possible to extract
only the products of quark-antiquark Boer-Mulders functions.
In order to separate them, in Refs.~\cite{Zhang:2008nu,Lu:2009ip} the positivity bound has been employed, obtaining an allowed range for each distribution. However 
one should add to this range the statistical errors of the fit,
which results in a very large final uncertainty
on the magnitude of $h_{1}^{\perp}$. 

The $pp$ or $p D$ 
$\cos 2 \phi$ DY asymmetry, which involves 
the sea distributions, is very small. 
A larger asymmetry 
is predicted for $p \bar p$ DY 
production 
\cite{Boer:2002ju,Bianconi:2004wu,Sissakian:2005yp,Gamberg:2005ip,Barone:2006ws}, a process to be studied 
in the next years at the GSI High-Energy Storage 
Ring \cite{Kotulla:2004ii,Lutz:2009ff,Barone:2005pu}. 
Another reaction probing valence distributions 
is $\pi N$ DY, under investigation by the COMPASS 
collaboration in their hadronic program \cite{Ketzer:2009zz}. 

A combined analysis of all these data will represent 
a decisive step towards a better knowledge of the 
Boer-Mulders function and of the phenomena 
originating from it.

\begin{acknowledgments}
We acknowledge support by the European Community - Research Infrastructure
Activity under the FP6 Program ``Structuring the European Research Area''
(HadronPhysics, contract number RII3-CT-2004-506078).
This work is also partially supported by the Helmholtz Association through
funds provided to the virtual institute ``Spin and Strong QCD''(VH-VI-231).
Authored by Jefferson Science Associates, LLC under U.S. DOE Contract 
No. DE-AC05-06OR23177.The U.S. Government retains a non-exclusive, 
paid-up, irrevocable, 
world-wide license to publish
or reproduce this manuscript for U.S. Government purposes.
\end{acknowledgments}



\end{document}